\documentclass[useAMS,usenatbib]{mn2e}
\usepackage{amssymb,amsmath,graphicx}

\def\gsim{\\raise 3pt\hbox{$\rangle$}\kern -8.5pt\raise -2pt\hbox{$\sim$}\ }
\def\lsim{\\raise 3pt\hbox{$\langle$}\kern -8.5pt\raise -2pt\hbox{$\sim$}\ }
\newtheorem{theorem}{Theorem}[section]
\newtheorem{corollary}[theorem]{Corollary}

\title[The Generalized Spectral Kurtosis Estimator]{The Generalized Spectral Kurtosis Estimator}
\author[G. M. Nita and D. E. Gary]{G. M. Nita$^{1}$\thanks{E-mail: gnita@njit.edu} and D. E. Gary$^{1}$\thanks{We acknowledge support for this work through NSF grant AST-0908344 to the New Jersey Institute of Technology.}\\
$^{1}$Center for Solar-Terrestrial Research, New Jersey
Institute of Technology, Newark, NJ 07102, USA}
\begin{document}
\date{}
\pagerange{\pageref{firstpage}--\pageref{lastpage}} \pubyear{2002}

\maketitle

\label{firstpage}

\begin{abstract}
Due to its conceptual simplicity and its proven effectiveness in real-time detection and removal of radio frequency interference (RFI) from radio astronomy data, the Spectral Kurtosis (SK) estimator is likely to become a standard tool of a new generation of radio telescopes. However, the SK estimator in its original form must be developed from instantaneous power spectral density (PSD) estimates, and hence cannot be employed as an RFI excision tool downstream of the data pipeline in existing instruments where any time averaging is performed. In this letter, we develop a generalized estimator with wider applicability for both instantaneous and averaged spectral data, which extends its practical use to a much larger pool of radio instruments.
\end{abstract}

\begin{keywords}
instrumentation: spectrographs(RFI)---methods: statistical(SK)
\end{keywords}

\section{Introduction}

The Spectral Kurtosis estimator ($\widehat{SK}$) was originally proposed by \citet{RFI1} as a statistical tool for real-time radio frequency interference (RFI) detection and excision in a Fast Fourier Transform (FFT) radio spectrograph. The first spectrograph designed for $\widehat{SK}$, the Korean Solar Radio Burst Locator \citep[KSRBL;][]{KSRBL}, demonstrated the effectiveness of the SK algorithm, but also revealed the need for a more accurate calculation of the theoretical RFI detection thresholds than initially proposed. Consequently, \citet{RFI2} derived the exact analytical expressions for the statistical moments of $\widehat{SK}$ and, based on its first four standard moments, assigned to it a Pearson Type IV probability curve \citep{Pearson}, which was shown to be in very good agreement with the Monte Carlo simulated $\widehat{SK}$ probability density function (pdf), as well as with the distribution derived from direct experimental observations made with the KSRBL instrument \citep{RFI3}.

As extensively described in the previous papers, what makes an SK spectrograph with $N$ spectral channels distinct from a traditional one is the fact that it accumulates not only a set of $M$ instantaneous power spectral density (PSD) estimates, denoted $S_1$, but also the squared spectral power denoted $S_2$. These sums, which have an implicit dependence on frequency channel $f_k$, are used to compute the averaged power spectrum $\langle{P}\rangle=S_1/M$, as well as the quantity
\begin{eqnarray}
\label{SK}
\widehat{SK}=\frac{M+1}{M-1}\Big{(}\frac{MS_2}{S_1^2}-1\Big{)},
\end{eqnarray}
which is a cumulant-based estimator of the spectral variability corresponding to the signal parent population,
\begin{eqnarray}
\label{Vk}
V_k^2=\frac{\sigma_k^2}{\mu_k^2},
\end{eqnarray}
where $\mu_k$ and $\sigma_k^2$ are the frequency-dependent PSD population means and variances, respectively. For a normally distributed time domain signal, i.e. an RFI-free signal, \citet{RFI2} showed that the estimator given by equation (\ref{SK}) is unbiased, i.e. $E(\widehat{SK})=V_k^2=1$.
However, this distinctive feature of an SK spectrograph prevents the employment of $\widehat{SK}$ as an RFI excision algorithm in an already existing instrument that is hardware limited to output only an averaged power spectrum, $\langle{P}\rangle$, without any possibility to intercept the instantaneous spectra needed to accumulate $S_2$ used in $\widehat{SK}$. To overcome this hardware limitation, we introduce in this letter a generalized SK estimator defined in terms of the sums $S_1=\Sigma\langle{P}\rangle_N$ and $S_2=\Sigma\langle{P}\rangle_N^2$, where  $\langle{P}\rangle_N=\Sigma_{i=1}^NP_i/N$ represents the averaged power spectrum over an arbitrary number $N$ of instantaneous  FFT spectra, which is the spectrograph output, while the outer sums are taken over $M$ such consecutive outputs. This new generalized $\widehat{SK}$ reduces to equation~(\ref{SK}) when $N=1$.
\section{A Generalization of the Spectral Kurtosis Estimator}
The generalization of the estimator $\widehat{SK}$ may be achieved directly from a fundamental property proven by \citet{RFI2}, which pertains to any gamma distribution
\begin{eqnarray}
\label{gamma}
f(x,a,d)=\frac{x^{d-1}e^{-\frac{x}{a}}}{a^d\Gamma(d)},
\end{eqnarray}
where $\Gamma(z)=\int_0^\infty{t}^{z-1}e^{-t}dt$ is the well known Euler's Gamma function.
This property, which is the basis of the whole SK concept, may be stated as follows
\begin{theorem}
\label{theorem}
Given a set of $M$ independent random variables $\{x_i\}$ sampled from a parent population described by a gamma distribution $f(x,a,d)$, the infinite set of statistical moments of the ratio $M S_2/S_1^2$, where $S_1=\Sigma_{i=1}^Mx_i$ and $S_2=\Sigma_{i=1}^Mx_i^2$, are given by the expression
\begin{eqnarray}
\label{ratio}
E\left[\left(\frac{MS_2}{S_1^2}\right)^n\right]&=&\frac{M^n\Gamma(Md)}{\Gamma(d)^M\Gamma(Md+2n)}\\\nonumber&&\times\frac{\partial^n}{\partial{t}^n}\Big{[}{\sum_{r=0}^n\frac{1}{r!}\Gamma(2r+d)t^r}\Big{]}^M\Big{|}_{t=0},
\end{eqnarray}
which is independent of the mean of the underlying distribution.
\end{theorem}
In \citep{RFI2} we proved this property in the particular cases $d=1$ (exponential distribution) and $d=1/2$ ($\chi^2$ distribution). However, following the same exact steps, it may be shown that Theorem \ref{theorem} generally holds for arbitrary values of $d$ \citep{PS}.

\citet{RFI2} used the particular forms of equation (\ref{ratio}) corresponding to $d=1$ to derive the $\widehat{SK}$ estimator given by equation(\ref{SK}), and to $d=1/2$ to derive a time domain kurtosis (TDK) estimator
\begin{eqnarray}
\label{TDK}
\widehat{K}=\frac{M+2}{M-1}\Big{(}\frac{MS_2}{S_1^2}-1\Big{)},
\end{eqnarray}
both of them being unbiased estimators of the spectral variability corresponding to the underlying probability distribution $f(x,a,d)$, which according to equation (\ref{Vk}) is
\begin{eqnarray}
\label{GVk}
V^2=E(x^2)/E(x)^2-1=1/d,
\end{eqnarray}
which is $1$ for $d=1$, and $2$ for $d=1/2$.

However, since the final goal of this study is to provide a generalized $\widehat{SK}$ that would work for arbitrary distribution functions $f(x,a,d)$, we find at this point useful to define it in a form that would provide an unbiased estimation of the normalized spectral variability $V^2d=1$, rather than $V^2$. This transformation gives the generalized estimator a fixed expectation for any $d$, while leaving its statistical properties unchanged, thus preserving its performance as a statistical detector of outliers.

Therefore, considering the result given by equation~(\ref{ratio}), we define the generalized $\widehat{SK}$ estimator as
\begin{eqnarray}
\label{GSK}
\widehat{SK}=\frac{Md+1}{M-1}\Big{(}\frac{MS_2}{S_1^2}-1\Big{)},
\end{eqnarray}
which, for any observable $x$ sampled from a gamma distribution $f(x,a,d)$, is an unbiased estimator of the normalized variability $V^2d$, i.e. $E(\widehat{SK})\equiv1$. Note that, for $d=1$, $\widehat{SK}$ reduces to the expression given by equation ({\ref{SK}), while for $d=1/2$, the original TDK estimator has to be modified according to equation (\ref{GSK}).

Getting back to the original motivation behind this study, i.e. the problem of adapting our original RFI excision algorithm to a spectrograph that is hardware constrained to output only power estimates already averaged over $N$ onboard accumulations, we consider the case of having as the only available observable the mean $\langle{x}\rangle_N=(1/N)\Sigma_{j=1}^Nx_i$. Since the probability distribution of the mean of $N$ independent random variables sampled from a gamma distribution is also a gamma distribution given by $f(\langle{x}\rangle_N,a/N,Nd)$, \citep{RFI2}, it immediately follows that the mean $\langle{x}\rangle$ satisfies the condition required by Theorem \ref{theorem}, and its associated $\widehat{SK}$ estimator may be defined according to equation (\ref{GSK}), where $S_1=\Sigma_{i=1}^M(\langle{x}\rangle_N)_i$, $S_2=\Sigma_{i=1}^M(\langle{x}\rangle_N)_i^2$, and $d\rightarrow{Nd}$. Moreover, since $\langle{x}\rangle_N=(1/N)\Sigma_{j=1}^Nx_j$, the sums entering equation (\ref{GSK}) may be simply replaced by the double sums $S_1=\Sigma_{i=1}^M(\Sigma_{j=1}^Nx_j)_i$ and $S_2=\Sigma_{i=1}^M(\Sigma_{j=1}^Nx_j)_i^2$.

Hence, we obtain the generalized definition of the Spectral Kurtosis estimator, which we state as follows:
\begin{corollary}
\label{corrolary}
Given a set of $M\times N$ independent random variables $\{x_i\}$ sampled from a parent population described by a gamma distribution $f(x,a,d)$, the expression
\begin{eqnarray}
\label{GSK_final}
\widehat{SK}=\frac{MN d+1}{M-1}\Big{(}\frac{MS_2}{S_1^2}-1\Big{)},
\end{eqnarray}
where $S_1=\Sigma_{i=1}^M(\Sigma_{j=1}^Nx_j)_i$ and $S_2=\Sigma_{i=1}^M(\Sigma_{j=1}^Nx_j)_i^2$,  is an unbiased estimator that, in the absence of any outlier contamination, is expected to evaluate to unity independently of the particular values of the parameters involved.
\end{corollary}
\section{The PDF of the Generalized Spectral Kurtosis Estimator}
Since the expectations $E[(\widehat{SK})^n]$ may be expressed in terms of the expectations $E[(MS_2/S_1^2)^n]$ given by equation (\ref{ratio}) amended by the substitution $d\rightarrow Nd$, all statistical moments of $\widehat{SK}$ are analytically defined, which in principle implies that its pdf may be uniquely determined from an infinite set of moments. However, it was experimentally shown by \citet{RFI3} that an approximation of the true $\widehat{SK}$ pdf based only on the first four statistical moments is sufficiently accurate to estimate the thresholds needed to flag RFI outliers with a predefined confidence level.  We follow the same approach here.

Using the standard notations $\mu_1'=E(\widehat{SK})\equiv1$ and $\mu_n=E[(\widehat{SK}-\mu_1')^n]$, the mean and first central moments of $\widehat{SK}$ are
\begin{eqnarray}
\label{GSKmom}
\mu_1'&=&1\\\nonumber\mu_2&=&\frac{2Nd(Nd+1)M^2\Gamma(MNd+2)}{(M-1)\Gamma(MNd+4)}\\\nonumber\mu_3&=&\frac{8Nd(Nd+1)M^3\Gamma(MNd+2)}{(M-1)^2\Gamma(MNd+6)}\\\nonumber&&\times((Nd+4)MNd-5Nd-2)\\\nonumber\mu_4&=&\frac{12Nd(Nd+1)M^4\Gamma(MNd+2)}{(M-1)^3\Gamma(MNd+8)}\\\nonumber&&\times(M^3N^4d^4+3M^2N^4d^4+M^3N^3d^3+68M^2N^3d^3\\\nonumber&&-93MN^3d^3+125M^2N^2d^2-245MN^2d^2\\\nonumber&&+84N^2d^2-32MNd+48Nd+24),
\end{eqnarray}
which also provide the standard statistical parameters $\beta_1=\mu_3^2/\mu_2^3$ and $\beta_2=\mu_4/\mu_2^2$ that are directly related to the more commonly used skewness, $\gamma_1=\sqrt{\beta_1}$ and kurtosis excess, $\gamma_2=\beta_2-3$.

A series expansion of its first standard moments,
\begin{eqnarray}
\label{exp}
\mu_2&\simeq&\frac{2}{M}\left[1+\frac{1}{d}\frac{1}{N}+O\left(\frac{1}{N^2}\right)\right]+O\left(\frac{1}{M^2}\right)\\\nonumber\gamma_1&\simeq&\frac{2\sqrt{2}}{\sqrt{M}}\left[1+\frac{7}{2d}\frac{1}{N}+O\left(\frac{1}{N^2}\right)\right]+O\left(\frac{1}{M^{3/2}}\right)\\\nonumber\gamma_2&\simeq&\frac{12}{M}\left[1+\frac{12}{d}\frac{1}{N}+O\left(\frac{1}{N^2}\right)\right]+O\left(\frac{1}{M^2}\right),
\end{eqnarray}
($O\left(x\right)$ is the Bachmann–-Landau notation meaning ``of order $x$'') is useful in assessing the influence on the shape of the $\widehat{SK}$ distribution due to large accumulation numbers $M$ and $N$. These expressions show that, although the inner accumulation over a large number of samples $N$ reduces the variance, skewness, and kurtosis excess, these parameters are lower bounded by the intrinsic limits $\mu_2=2/M$, $\gamma_1=2\sqrt{2}/\sqrt{M}$, and $\gamma_2=12/M$ dictated by the outer accumulation number $M$. Particularly, for any $N$, the skewness of $\widehat{SK}$ does not vanish faster than $2\sqrt{2}/\sqrt{M}$, which indicates that the asymmetry of its pdf should be considered in calculating the RFI thresholds even for fairly large accumulation numbers $M$. Moreover, equation~(\ref{exp}) also indicates that for excessive averaging, i.e. large $N$, the shape of the $\widehat{SK}$ pdf becomes practically independent of $d$, which means that the estimator loses its ability to distinguish outliers that would belong to a gamma, or gamma-like distribution characterized by a different parameter $d$.  Thus, the performance of $\widehat{SK}$ is improved by limiting $N$, optimally to $N=1$.

To find the appropriate pdf approximation type, one has to investigate the behavior of the Pearson criterion \citep{Pearson,Kendall}
\begin{eqnarray}
\label{criterion}
\kappa=\frac{\beta_1(\beta_2+3)^2}{4(4\beta_2-3\beta_1)(2\beta_2-3\beta_1-6)}.
\end{eqnarray}
Figure~\ref{k} shows the $\kappa$ diagram obtained for different combinations of the accumulation number $M$ and the product $Nd$. The diagram shows two main regions labeled Type IV ($0<\kappa<1$) and Type III ($\kappa>1$), as well as a narrow region (black) that corresponds to a Pearson Type I ($\kappa<0$) pdf. The horizontal dotted line indicate the values of $N$ (14 for $d=1$ and 28 for $d=1/2$) above which it can be analytically proven, by evaluating the limit of $\kappa$ for $M\rightarrow\infty$, that the Person Type IV condition is not satisfied for any value of $M$.
\begin{figure}
  \includegraphics[width=80mm]{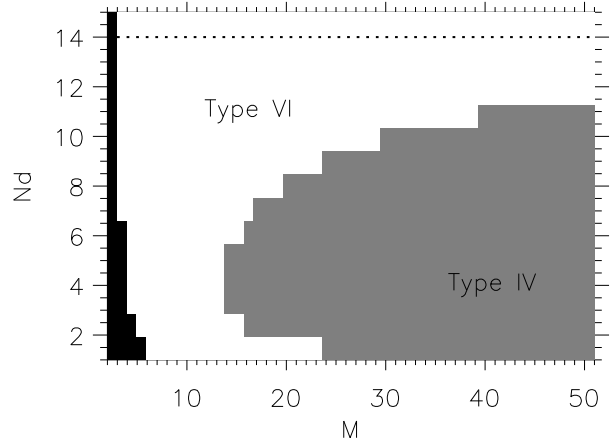}
  \caption{\label{k}The $\kappa$ diagram obtained for different combinations of $M$ and $N d$. The black region corresponds to $\kappa<0$ (Type I), the gray region to $0<\kappa<1$ (Type IV), and the white region to $\kappa>1$ (Type VI).
  The horizontal dotted line indicate the values of $N d$ above which the condition required by a Pearson Type IV pdf is not satisfied for any value of $M$.
  }
\end{figure}
Since the case of the Type IV pdf, which exactly matches the first four moments of the true $\widehat{SK}$ pdf, was extensively addressed by \citet{RFI2}, we refer the reader to our previous work for details regarding the use of this particular type for computing the RFI detection thresholds. We consider the narrow region corresponding to a Type I approximation to be out of the range of interest for practical applications, but refer the interested reader to \citet{Kendall} for a detailed analysis connecting the moments of the true distribution to parameters defining the Type I pdf. Nevertheless, since the parameter region satisfying the conditions required by the Type VI pdf is the most relevant in connection to the generalization of the $\widehat{SK}$ estimator, we address it in the next section. By doing so we also extend the study of the statistical properties of $\widehat{SK}$ for $N=1$ to $5/d<M<23/d$, which was not addressed in our original study.
\subsection{The Pearson Type VI PDF}
The Pearson Type VI pdf is a beta distribution of second kind \citep{Kendall}, which is commonly defined on the $[0,\infty)$ interval as
\begin{eqnarray}
\label{type6}
p(x,\alpha,\beta)=\frac{\Gamma(\alpha+\beta)}{\Gamma(\alpha)\Gamma(\beta)}\frac{x^{\alpha-1}  }{(1+x)^{\alpha+\beta}},
\end{eqnarray}
where $\alpha$ and $\beta$ are two adjustable parameters that determine its shape. If  $\Re(\alpha)>0$ and $\Re(\beta)>n>0$, the distribution satisfies the normalization condition $\int_0^\infty p(x,\alpha,\beta)dx=1$, and its mean and central moments to order $n$ exist. Under these minimal conditions, the mean and central moments to order $4$ of the beta distribution are
\begin{eqnarray}
\label{type6mom}
\mu_1'&=&\frac{\alpha}{\beta-1};\;\mu_2=\frac{\alpha(\alpha+\beta-1)}{(\beta-2)(\beta-1)^2}\\\nonumber\mu_3&=&\frac{2\alpha(\alpha+\beta-1)(2\alpha+\beta-1)}{(\beta-1)^3((\beta-5)\beta+6)}\\\nonumber\mu_4&=&\frac{3\alpha(\alpha+\beta-1)}{(\beta-4)(\beta-3)(\beta-2)(\beta-1)^4}\\\nonumber&&\times\left[(\beta+5)\alpha^2+(\beta-1)(\beta+5)\alpha+2(\beta-1)^2\right].
\end{eqnarray}
In order to reproduce the shape of the $\widehat{SK}$ pdf, we equate the second and the third central moments given by equation (\ref{type6mom}) with the corresponding moments of the $\widehat{SK}$ pdf given by equation (\ref{GSKmom}) to obtain
\begin{eqnarray}
\label{type6sol}
\alpha&=&\frac{1}{\mu_3^3}\Big[32\mu_2^5-4\mu_3\mu_2^3+8\mu_3^2\mu_2^2+\mu_3^2\mu_2-\mu_3^3\\\nonumber&&+\left(8\mu_2^3-\mu_3\mu_2+\mu_3^2\right)\sqrt{16\mu_2^4+4\mu_3^2\mu_2+\mu_3^2}\Big]\\\nonumber\beta&=&3+\frac{2\mu_2}{\mu_3^2}\left[4\mu_2^2+\sqrt{16\mu_2^4+4\mu_3^2\mu_2+\mu_3^2}\right].
\end{eqnarray}
The condition $\mu_1'=1$ can be then satisfied by introducing the translation parameter
$\delta=(\beta-\alpha-1)/(\beta-1)$,
which changes the support of the beta distribution to $[\delta,\infty)$, without changing its central moments.

Therefore, the distribution $p(x-\delta,\alpha,\beta)$ exactly matches the mean, variance, and kurtosis of the $\widehat{SK}$ distribution, and the RFI probability of false alarm (pfa) corresponding to a threshold located at the ordinate $\xi$ can be estimated by using the cumulative function $\texttt{CF}(\xi)=\int_0^{\xi-\delta}p(x,\alpha,\beta)dx$ or, alternatively, the complementary cumulative function, $\texttt{CCF}(\xi)=\int_{\xi-\delta}^{\infty}p(x,\alpha,\beta)dx$. These probabilities are given by
\begin{eqnarray}
\label{CFtype6}
\texttt{CF}(\xi)=\frac{\Gamma(\alpha+\beta)}{\Gamma(\beta)}(\xi-\delta)^{\alpha}\tilde{F}(\alpha,\alpha+\beta,\alpha+1,\delta-\xi )\\\nonumber
\end{eqnarray}
and
\begin{eqnarray}
\label{CCFtype6}
\texttt{CCF}(\xi)=\frac{\Gamma(\alpha+\beta)}{\Gamma(\alpha)}(\xi-\delta)^{-\beta}\tilde{F}\left(\beta,\alpha+\beta,\beta+1,\frac{1}{\delta-\xi}\right),
\end{eqnarray}
where
\begin{eqnarray}\nonumber
\tilde{F}(a,b,c,z)=\frac{1}{\Gamma(a)\Gamma(b)}\sum_{n=0}^\infty\frac{\Gamma(a+n)\Gamma(b+n)}{\Gamma(c+n)}\frac{z^n}{n!}
\end{eqnarray}
is the regularized Gauss hypergeometric series, which is absolutely convergent for finite parameters $a$, $b$ and $c$, if $|z|<1$, conditions that are automatically satisfied either by $\xi-\delta$ or by $1/(\xi-\delta)$, which enter as hypergeometric function arguments in equations ({\ref{CFtype6}) and (\ref{CCFtype6}), respectively. However, as previously shown in the case of the Type IV pdf \citep{RFI2}, which also requires hypergeometric series evaluations, to compute the pfa with sufficient accuracy for practical applications, a simple numerical integration of the $\widehat{SK}$ pdf may be considered as a viable alternative to using the analytical expressions given by equations ({\ref{CFtype6},\ref{CCFtype6}).
\subsection{Accuracy of the Pearson Type VI approximation}
Since the Type VI approximation exactly matches only the mean, variance and kurtosis of the true $\widehat{SK}$ pdf, we test the accuracy of this approximation by evaluating the error of the fourth moment $\mu_4$ provided by equation (\ref{type6mom}) relative to the exact moment provided by equation (\ref{GSKmom}). We display in Figure \ref{type6err} a set of contour lines indicating, as absolute values, the relative error levels of $0.1\%$ and $0.5\%$ obtained over an extended range spanned by $M$ and $Nd$. The gray shaded region corresponds to the range of $(M,N,d)$ parameters that allows a Pearson Type IV approximation ($\kappa<1$). However, the Type VI relative error contours also extend over this region just because, while the condition ($\kappa>1$) prevents a four--moment Type IV approximation in the white region, the ($\kappa<1$) condition still allows a three--moment Type VI approximation in the gray shaded region.
\begin{figure}
\includegraphics[width=80mm]{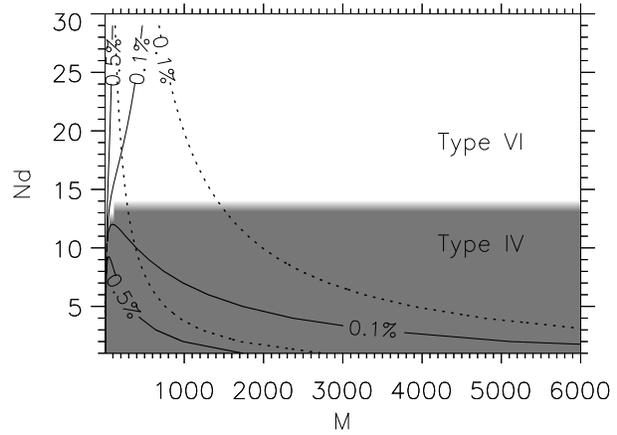}\caption{\label{type6err}The $0.1\%$ and  $0.5\%$ contours of the error (absolute values) of the fourth central moment of the Type VI pdf approximation relative to the exact moment given by equation (\ref{GSKmom}) are shown by solid lines. The fourth moment error contours (absolute values) corresponding to the Type III approximation are shown by dotted lines.
}
\end{figure}
In our opinion, the result of this analysis is remarkable not only due to the fact the Type VI approximation reproduces the fourth moment of the true $\widehat{SK}$ distribution with a relative error much smaller than $0.1\%$ over most of its parameter range, but also, starting with a not too large accumulation number $N$, in the region in which a Pearson Type IV approximation is possible.
Therefore, guided by the practical goal of estimating with reasonable accuracy the tail probabilities of the $\widehat{SK}$ pdf, we seek a more convenient approximation to the $\widehat{SK}$ pdf that, over the range of interest for practical applications, would provide similar accuracy to the Type IV or Type VI approximations.
A good candidate is the Type III pdf that, although strictly corresponding to an infinite value of the criterion $\kappa$, may provide an accurate approximation in the case of large positive values of $\kappa$ that results from small values in the denominator of equation(\ref{criterion}). The motivation behind this investigation is that the Type III pdf is the more convenient gamma distribution, also able to provide a three---moment based approximation \citep{Pearson}.

\subsection{The Pearson Type III approximation of the SK distribution}
As in the case of the Type VI pdf, in order to match the mean $\mu_1'=1$ of the $\widehat{SK}$ distribution, we introduce the location parameter $\delta$ and define the Type III pdf as,
\begin{eqnarray}
p(x,\alpha,\beta,\delta)=\frac{(x-\delta)^{\beta-1}e^{-\frac{x-\delta}{\alpha}}}{a^\beta\Gamma(\beta)},
\end{eqnarray}
which is the gamma distribution $f(x-\delta,\alpha,\beta)$ having the first four moments given by
\begin{eqnarray}
\label{type3mom}
\mu_1'=\alpha\beta+\delta;\;\mu_2=\alpha^2\beta;\;\mu_3=2\alpha^3\beta;\;\mu_4=3\alpha^4\beta(\beta+2).
\end{eqnarray}

Hence, the first three relationships provide
\begin{eqnarray}
\label{type3sol}
\alpha=\frac{\mu _3}{2\mu _2};\;\beta=\frac{4\mu _2^3}{\mu _3^2};\;\delta=1-\frac{2\mu _2^2}{\mu _3},
\end{eqnarray}
and replacing $\mu_2$ and $\mu_3$ by the expressions (\ref{GSKmom}) we complete the solution of a Pearson Type III approximation for the $\widehat{SK}$ pdf.

The Pearson Type III CF is
\begin{eqnarray}
\label{type3cf}
CF(\xi,\alpha,\beta,\delta)=\Gamma_x\left(\beta,\frac{\xi -\delta }{\alpha}\right)\Big/\Gamma(\beta),
\end{eqnarray}
where $\Gamma_x(\beta,x)=\int_0^x t^{\beta-1}e^{-t}dt$ is the incomplete gamma function.

To investigate the accuracy of this approximation, we compute the relative error of the fourth moment given by equation (\ref{type3mom}), relative to the exact moment of the $\widehat{SK}$ distribution. Using equation (\ref{type3sol}), this error can be expressed
\begin{eqnarray}
\label{type3error}
\epsilon_{\mu_4}=\mu_4^{III}/\mu_4^{\widehat{SK}}-1=\left(3\beta_1-2\beta_2+6\right)/\left(2\beta_2\right),
\end{eqnarray}
which compared with the criterion $\kappa$ given by (\ref{k}) shows that, indeed, while  a perfect match of $\mu_4^{\widehat{SK}}$ would result in an infinite value of $\kappa$, the Type III approximation may still reproduce the $\widehat{SK}$ distribution with reasonable accuracy if a favorable combination of parameters is met.

The dotted lines in Figure~\ref{type6err} show, as absolute values, the Type III error contours for $Nd\in[1,30]$ and $M\in [2,6000]$. This figure suggests that the three-moment Type III approximation might be accurate enough to allow a uniform approach over most of the parameter space relevant for practical applications, with the benefit of allowing the use of the more convenient equation (\ref{type3cf}) for estimating the pfa of the $\widehat{SK}$ pdf.

As a more quantitative argument in support of this assertion, we present in Table \ref{test} the threshold solutions obtained for $d=1$, $N=10$, and $M=300$, based on the numerical evaluation of the analytical expressions of the CF and CCF corresponding to the Type IV \citep[equation (61),][]{RFI2}, Type VI (equations [\ref{CFtype6},\ref{CCFtype6}]) and Type III (equation [\ref{type3cf}]) approximations. The first two columns display the lower and upper thresholds, the third column displays the error in estimating the exact $\mu_4$ moment, while the last two columns display the difference between the Type IV pfa corresponding to the computed thresholds and the predefined target pfa, which was chosen to be $0.13499\%$ at both ends of the distribution, the  same as a $3\sigma$ standard error. For comparison, we show in the last row the results corresponding to the symmetrical thresholds that result from a normal pdf approximation.

These results show that in this particular case, which corresponds to $\epsilon_{\mu_4}<1\%$, the supplemental data loss that results from adopting the thresholds estimated based on the Type III approximation does not exceed the Type IV pfa (the best available estimate) by more than $0.01\%$. Therefore, based on the results shown in Figure \ref{type6err}, we conclude that the Type III approximation would also provide similar accuracy over most of the range of interest for practical applications. On the other hand, we point out that ignoring the intrinsic skewness of the $\widehat{SK}$ estimator, the symmetric thresholds of $1\pm3\sqrt{\mu_2}$ would result in more significant data loss, as well as in a reduced performance in detecting certain types of RFI, as confirmed by the experimental results reported by \citet{RFI3}.
\begin{table}
\caption{Threshold solutions obtained from different pdf approximations for $N=10$, $M=300$, and $d=1$\label{test}}
\begin{tabular}{lrrrrr}
\hline
Method&Lower&Upper&$\epsilon_{\mu_4}(\%)$&$\delta_{pfa}^{Lower}(\%)$&$\delta_{pfa}^{Upper}(\%)$\\
\hline
Type IV&0.76613&1.28345&0&0&0\\Type VI&0.76648&1.28313&0.18&0.002&0.001\\Type III&0.76754&1.28212&-0.71&0.009&0.006\\Normal&0.74288&1.25712&-100&-0.093&0.162\\
\hline
\end{tabular}
\end{table}

We have tested the algorithm on both simulated data and solar observations (from KSRBL), and find excellent agreement with the theoretical pdf for the RFI-free case, with various choices of $M$ and $N$.  Space limitations in this Letter preclude detailed discussion.  However, we note that the use of GSK places significant constraints on the instrumental stability over the outer accumulation $M,$ since the theory assumes stationarity in the statistical behavior of the Gaussian noise.  Even subtle fluctuations in gain, for example, cause a shift of the pdf mean above unity, resulting in excess flagging of RFI-free data (i.e. increase in pfa).  For instruments with a long integration time $\tau$, the product $M\tau$ can become quite long (seconds), during which the system temperature ($T_{\rm sys}$) and gain must be constant for RFI-free data channels.  Of course, $T_{\rm sys}$ variations in frequency, and on time scales long compared to $M\tau$, are permitted.  \citet{RFI1} discuss the effects and limitations of accumulating during times of changing $T_{\rm sys}$.

\section{Conclusions}
We have defined a generalized Spectral Kurtosis estimator that allows the option of using already averaged spectra as input for the SK--based RFI excision algorithm originally proposed by \citet{RFI1}. We derived the exact analytical expressions providing its infinite set of statistical moments and we identified two main regions of the parameter space in which  $\widehat{SK}$ pdf may be best approximated by a Pearson Type IV or Type VI pdf. However, we showed that a simple three-moment Pearson Type III approximation may be accurate enough over most of the parameter space to justify its use as a less computationally demanding alternative to the other approximations investigated. The generalized $\widehat{SK}$ is applicable to either instantaneous or averaged data, and has the advantage of allowing its straightforward implementation as a RFI detection and excision tool in the data pipeline of already existing instruments, independently of the method used to obtain the PSD estimates, whether from FFT or narrow band time domain power detectors.

Use of the generalized $\widehat{SK}$ is likely to be superior to other means of detecting RFI in averaged data based on simple power-level thresholds because it is always 1 for RFI-free data, and hence eliminates the need to determine the mean background power level.  However, the performance of $\widehat{SK}$ is improved by limiting $N$, and newly designed instruments should optimally accumulate sums of power and power-squared on instantaneous spectra (i.e. $N=1$), as described in \citet{RFI3}.

\bsp
\label{lastpage}

\end{document}